\documentclass[
    aps,prb,10pt,
    twocolumn,reprint,
    superscriptaddress,floatfix,amsmath,amssymb
]{revtex4-2}

\usepackage{graphicx}
\graphicspath{{}}
\usepackage[caption=false]{subfig}

\usepackage{dcolumn}
\usepackage{bm}
\usepackage{hyperref}

\usepackage{color}
\usepackage{siunitx}
\usepackage{algorithm}
\usepackage{algpseudocode}
\usepackage{float}
\usepackage{braket}
\usepackage[export]{adjustbox}
\usepackage{placeins}

\begin{document}

\title{Magnetic phases of the anisotropic triangular lattice Hubbard model}

\def\umphys{%
    Department of Physics, University of Michigan,
    Ann Arbor, MI 48109, USA
}%

\def\oakridge{%
    Materials Science and Technology Division, Oak Ridge National Laboratory,
    Oak Ridge, TN 37831, USA
}%

\author{Yang Yu}
\affiliation{\umphys}

\author{Shaozhi Li}
\affiliation{\oakridge}

\author{Sergei Iskakov}
\affiliation{\umphys}

\author{Emanuel Gull}
\affiliation{\umphys}

\date{\today}

\begin{abstract}
The Hubbard model on an anisotropic triangular lattice in two dimensions, a fundamental model for frustrated electron physics, displays a wide variety of phases and phase transitions. 
This work investigates the model using the ladder dual fermion approximation which captures local correlations non-perturbatively but approximates non-local correlations. 
We find metallic, one-dimensional antiferromagnetic, non-collinear antiferromagnetic, square-lattice antiferromagnetic, and spiral phases but no evidence of collinear antiferromagnetic order in different parts of the phase diagram. Analyzing the spin susceptibility in detail, we see both regions of agreement and of discrepancy with previous work. The case of Cs$_2$CuCl$_4$ is discussed in detail.  
\end{abstract}

\maketitle

\section{Introduction}

Correlation physics on triangular lattices is a subject of intense interest. Many triangular lattice magnets including Cs$_2$CuCl$_4$~\cite{coldea2003}, $\kappa$-(BEDT-TTF)$_2$Cu$_2$(CN)$_3$~\cite{shimizu2003}, EtMe$_3$Sb[Pd(dmit)$_2$]$_2$~\cite{itou2008},  NiGa$_2$S$_4$~\cite{nakatsuji2005}, Ba$_3$CuSb$_2$O$_9$~\cite{zhou2011}, Ba$_3$CoSb$_2$O$_9$~\cite{shirata2012}, YbMgGaO$_4$~\cite{li2015}, and NaYbO$_2$~\cite{bordelon2019} have been experimentally investigated. Transition metal dichalcogenide moiré superlattices~\cite{wu2018,zhang2021,regan2022} and triangular optical lattices~\cite{struck2011,yang2021} have also recently been proposed as platforms to study correlation physics on the triangular lattice. Theoretically, both triangular lattice spin models and the triangular lattice Hubbard model have been investigated to describe frustrated quantum magnets. Nevertheless, the phase diagrams of triangular lattice systems are still under debate both experimentally~\cite{xu2016,ni2019,bourgeois-hope2019} and theoretically~\cite{savary2016,zhou2017}. Quantum spin liquid (QSL) states~\cite{anderson1973,anderson1987,balents2010,knolle2019,broholm2020} and Mott transitions~\cite{lee2006, kanoda2011,powell2011,furukawa2015} proposed in triangular lattice systems therefore remain elusive.

The case of spatially anisotropic hopping/exchange interaction is particularly interesting. Previous studies show inconsistencies for both the anisotropic triangular lattice Heisenberg and Hubbard model. Among those is the existence of collinear antiferromagnetic order in the limit of weakly coupled one-dimensional chains~\cite{weng2006,hayashi2007,hauke2011,hauke2013,reuther2011,yunoki2006,heidarian2009,ghorbani2016,castells-graells2019,gonzalez2020,
starykh2007,starykh2010,pardini2008,bishop2009,ghamari2011,thesberg2014,
yamada2014a,tocchio2014,szasz2021,
yamada2014,acheche2016}.

Here we present a dual fermion (DF) study of the anisotropic triangular lattice Hubbard model. The DF method, as one of the diagrammatic extensions of the dynamical mean field theory (DMFT), is unbiased (in the sense that it does not favor certain types of magnetic orders) and provides high-resolution spin susceptibilities, which renders it a valuable tool in discussing the above-mentioned problem. We survey the phase diagram, connect it to ground state simulations, and discuss in detail the parameter believed to be relevant for Cs$_2$CuCl$_4$.

The remainder of this paper proceeds as follows. In Sec.~\ref{sec:method}, we give a brief introduction to the methodology by describing the model, the dual fermion method, the observables of interest, and the possible magnetic orders. In Sec.~\ref{sec:results}, we present our phase diagram, discuss the phases and phase transitions, and apply the method to the concrete example of Cs$_2$CuCl$_4$. Conclusions are drawn in Sec.~\ref{sec:conclusion}.

\section{Method}\label{sec:method}

\subsection{Hubbard model}
We study the  Hubbard model \cite{arovas2022,qin2022}
\begin{equation}
    \hat{H}=-\sum_{\langle i j\rangle\sigma} t_{i j} \hat{c}_{i \sigma}^{\dagger} \hat{c}_{j \sigma}+\text{H.c.}+U \sum_{i} \hat{n}_{i \uparrow} \hat{n}_{i \downarrow}-\mu \sum_{i\sigma} \hat{n}_{i \sigma}
    \label{eqn:Hubbard_model}
\end{equation}
on an anisotropic triangular lattice in two dimensions at half filling.
Here $\hat{c}_{i \sigma}$ ($\hat{c}_{i \sigma}^{\dagger}$) are fermion annihilation (creation) operators on site $i$ with spin $\sigma=\uparrow,\downarrow$. $\hat{n}_{i \sigma}=\hat{c}_{i \sigma}^{\dagger}\hat{c}_{i \sigma}$ is the particle number operator. $\langle i j\rangle$  denotes nearest-neighbor pairs on the triangular lattice. $U$ is the on-site interaction.  The anisotropic hopping is set to be $t_{ij}=t$ for two of the three directions and $t_{ij}=t^{\prime}$ for the third direction as illustrated in Fig.~\ref{fig:magnetic_order}(c). When $t^{\prime} \gg t$, the system approaches the limit of decoupled one-dimensional chains, and when $t^{\prime} \ll t$, the system becomes a square lattice. The chemical potential $\mu$ is chosen such that the system is close to half filling (we adjust $\mu$ such that $|\sum_{\sigma}\langle \hat{n}_{i \sigma}\rangle-1|<0.01$). All the data in this paper are calculated at fixed temperature $T=0.1t$ ($\beta=1/T=10/t$) unless stated otherwise. In the remainder of the paper, the hopping $t$ is set to $1$ and used as an overall energy scale.

\subsection{Dual fermion expansion and ladder dual fermion approximation}

The DF expansion \cite{rubtsov2008} is a diagrammatic technique developed to describe non-local correlation effects in models with local interactions. It can be understood as an expansion around a local limit in terms of `dual' diagrams. At the lowest order, only local correlations are captured, but the method converges to the exact limit when all `dual' diagrams to all orders are summed. In contrast to other diagrammatic techniques, it provides both a non-perturbative strongly correlated local self-energy and non-local contributions on a continuous momentum grid, as well as direct access to one- and two-particle correlation functions.  A detailed discussion of the technique is presented in Refs.~\cite{rubtsov2008,brener2008,rubtsov2009,hafermann2009} and in a review~\cite{rohringer2018}.

In this paper, we employ the ladder dual fermion approximation (LDFA) with a dynamical mean field theory \cite{georges1996} starting point (see Refs.~\cite{lee2008,antipov2011,li2014,yudin2014,laubach2015,li2020} for previous DF work on the triangular lattice Hubbard model). This approximation only contains dual  particle-hole ladder diagrams, which are summed to infinite order. All non-ladder diagrams and all higher-order vertices are neglected. The quality of the approximation is assessed in Refs.~\cite{katanin2013,simonscollaborationonthemany-electronproblem2015,iskakov2016,ribic2017,leblanc2019} (see also \cite{gukelberger2017}). We use the implementation provided by the open-source \texttt{OpenDF} code \cite{antipov2015} based on \texttt{ALPSCore}~\cite{wallerberger2018,gaenko2017}.

As a starting point, the self-consistent vertices and self-energies of a DMFT impurity problem are used as the input of the LDFA~\cite{rohringer2012}. These quantities are obtained using numerically exact continuous-time auxiliary field quantum Monte Carlo~\cite{gull2008,gull2011,gull2011a,chen2015} (for analytic expressions of the non-interacting triangular lattice local density of states see Appendix~\ref{app:DOS}). The LDFA method proceeds by solving the Bethe-Salpeter equation (BSE) to obtain dual vertex functions. Once the dual vertex functions are obtained, the Schwinger Dyson equation (SDE) and the Dyson equation are used to obtain dual self-energies, which are used to update the dual Green's functions. The  BSE, SDE, and the Dyson equation are iterated until convergence~\cite{antipov2015}. This iterative procedure method may encounter convergence issues where the DMFT starting point is far from the exact solution, in particular in the vicinity of phase transitions and at low temperatures. 

LDFA results, such as one-body Green's function $G$ and spin susceptibility $\chi_{\text{s}}$, are obtained as a function of Matsubara frequencies. Dynamical quantities, such as spectral functions and dynamical spin susceptibilities, require analytic continuation from the Matsubara to the real axis. This procedure is ill-conditioned and uncontrolled \cite{jarrell1996}. In this paper, where possible, we therefore present quantities that can be directly extracted from the Matsubara results. 
Where we show analytically continued quantities, we use the maximum entropy method  implemented in the open-source \texttt{MaxEnt} code \cite{levy2017} based on \texttt{ALPSCore}~\cite{wallerberger2018,gaenko2017}.

Despite being approximate, the LDFA has advantages that make it a unique tool for studying correlated phases. First, since it contains strong local correlations non-perturbatively, it can be used to study strongly correlated systems outside the perturbative regime. Second, since it gives access to generalized susceptibilities with almost continuous momentum resolution (we use systems of size $24\times24$ and $48\times48$ in this work), it can be used to detect the emergence of phases with long-range and incommensurate orders. Finally, since the formalism is based on detecting large fluctuations, rather than entering an ordered state,  all possible orders are treated on equal footing without an \textit{a priori} assumption of an expected phase.

\subsection{Spin structure factor}

The main quantities of interest in this paper are  the dynamical spin susceptibilities and the dynamical spin structure factor.
The magnetic scattering cross section is directly related to the dynamical spin structure factor, 
\begin{equation}
    \mathcal{S}^{\alpha \alpha}(\boldsymbol{q}, \omega)
    =\frac{1}{N} \int_{-\infty}^{\infty} \frac{\mathrm{d} t}{2 \pi} \mathrm{e}^{\mathrm{i} \omega t}\left\langle \hat{S}_{\boldsymbol{q}}^{\alpha}(t) \hat{S}_{-\boldsymbol{q}}^{\alpha}(0)\right\rangle,
\end{equation}
where $ \hat{S}^{\alpha}_{\boldsymbol{q}}=\sum_{i} \mathrm{e}^{-\mathrm{i}\boldsymbol{q}\cdot \boldsymbol{r}_i}\hat{S}^{\alpha}_i$ denotes the Fourier transform of the spin operator with $\alpha=x,y,z$, and $N$ is the number of lattice sites. The total scattering cross section that integrates over all frequencies is related to the equal-time spin structure factor,
\begin{equation}
    \mathcal{S}^{\alpha \alpha}_{0}(\boldsymbol{q})
    = \frac{1}{N}\left\langle \hat{S}_{\boldsymbol{q}}^{\alpha}(0) \hat{S}_{-\boldsymbol{q}}^{\alpha}(0)\right\rangle.
    \label{eqn:ssf}
\end{equation}

Our calculation is performed in the normal state, where SU(2) symmetry is preserved, which allow us to focus on the $z$ component. 
The frequency- and momentum-dependent
dynamical spin susceptibility,
\begin{equation}
    \chi_{\text{s}}(\boldsymbol{q},\omega)
     =- \frac{\mathrm{i}}{N}  \int_{0}^{\infty} \mathrm{d} t  \mathrm{e}^{\mathrm{i} (\omega+\mathrm{i} 0^{+}) t}  \left\langle \left[\hat{S}^{z}_{\boldsymbol{q}}(t), \hat{S}^{z}_{-\boldsymbol{q}}(0)\right]\right\rangle,                                                         
\end{equation} 
which describes the response to a weak externally applied magnetic perturbation, can be extracted from the spin susceptibility on the Matsubara axis,
\begin{equation}
        \chi_{\text{s}}(\boldsymbol{q},\omega_n)  = -\frac{1}{N} \int_{0}^{\beta} \mathrm{d} \tau \mathrm{e}^{\mathrm{i} \omega_n \tau}\langle T_{\tau} \hat{S}^{z}_{\boldsymbol{q}}(\tau) \hat{S}^{z}_{-\boldsymbol{q}}(0)\rangle,
\end{equation}
via analytic continuation. (Note a minus sign difference with respective to the physical spin susceptibilities.)
The dynamical spin structure factor can then be written as
\begin{equation}
    \mathcal{S}^{z z}(\boldsymbol{q}, \omega)=-\frac{1}{\pi} \left[1+n_B(\omega)\right]  \operatorname{Im} \chi_{\text{s}}(\boldsymbol{q},\omega)
\end{equation}
via the fluctuation-dissipation relation~\cite{mahan2000a,sachdev2011}, where $n_B(\omega)=1/\left(\mathrm{e}^{\beta \omega}-1\right)$ denotes the Bose-Einstein distribution. The equal-time spin structure factor can be calculated from a sum over the Matsubara susceptibility as
\begin{equation}
    \mathcal{S}^{z z}_{0}(\boldsymbol{q})= - \frac{1}{\beta} \sum_{\omega_n} \chi_{\text{s}}(\boldsymbol{q},\omega_n). 
\end{equation}

\subsection{Magnetic orders}\label{subsection:magnetic_orders}

\begin{figure*}[htb]
    \centering
    \includegraphics[width=1\linewidth]{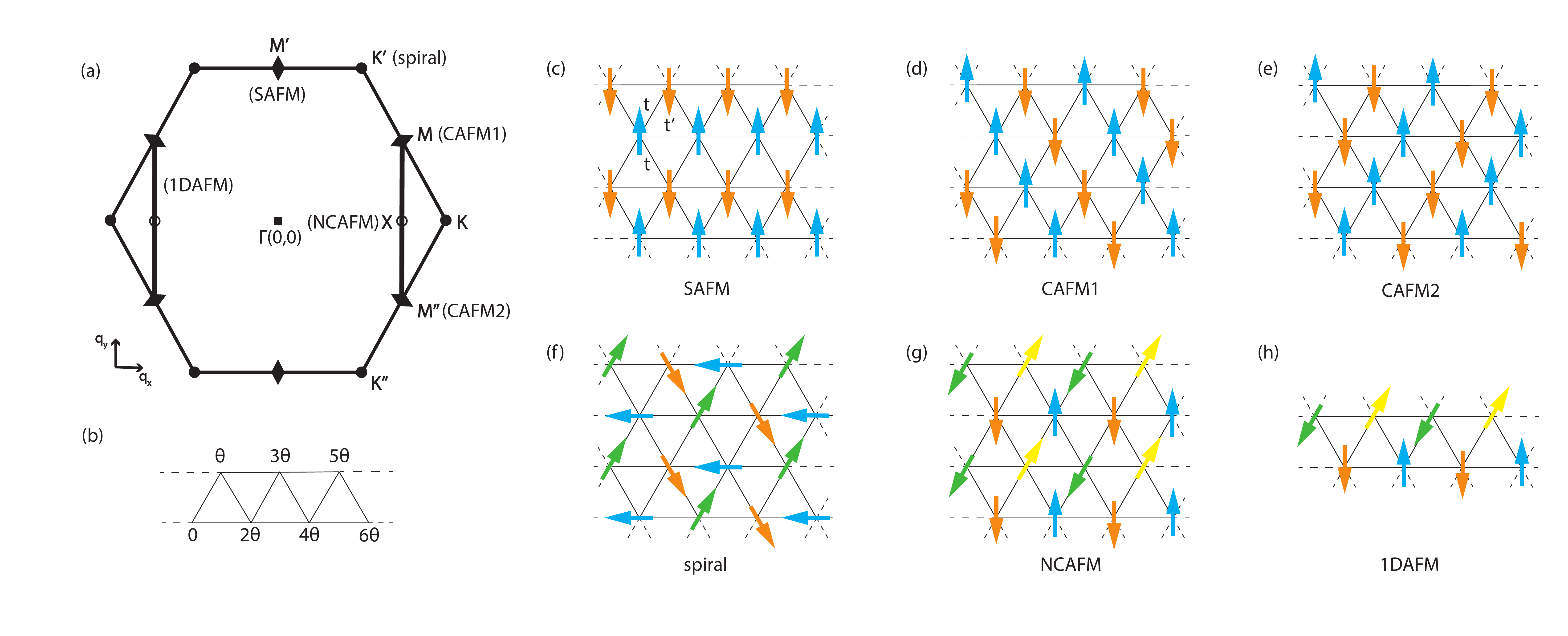}
    \caption{Ordering wave vectors for magnetic orders on the triangular lattice are represented by the different symbols or lines in (a) where the $\Gamma$ point is set to be the origin and the hexagon indicates the first Brillouin zone with side length equaling $4\pi/3$. The corresponding spin configurations in the real space are illustrated in: (c) square-lattice antiferromagnetic (SAFM), (d) and (e) collinear antiferromagnetic (CAFM1, CAFM2), (f) spiral, (g) non-collinear antiferromagnetic (NCAFM), and (h) one-dimensional antiferromagnetic (1DAFM) orders. The ground state of the classical Heisenberg model is illustrated in (b) where the relative angle between the spin on each site and the spin on the left-bottom site is indicated.}
    \label{fig:magnetic_order}
\end{figure*}

Upon approaching a magnetically ordered state, a divergence in the equal-time spin structure factor with the ordering wave vectors indicates the onset of the corresponding magnetic order. Since the LDFA calculation performed here always remains in the normal state, large well-defined peaks in the equal-time spin structure factor indicate strong magnetic correlations. These are viewed as a precursor of magnetic order. To compare our results with previous studies of related models at zero temperature, we identify strong magnetic correlations at the lowest temperature studied ($T=0.1$) with the corresponding magnetic orders at zero temperature in this work.

For the symmetry-broken magnetically ordered states, we can directly evaluate the ordering wave vectors from the definition in Eq.~(\ref{eqn:ssf}).
Figs.~\ref{fig:magnetic_order}(c)-~\ref{fig:magnetic_order}(h) show the real-space spin configurations for the magnetic orders discussed in this paper. The corresponding ordering wave vectors and equivalent points/lines are indicated by the different symbols in Fig.~\ref{fig:magnetic_order}(a). Square-lattice antiferromagnetic (SAFM) order (Fig.~\ref{fig:magnetic_order}(c)) with wave vector at M$^{\prime}$$(0,2\pi/\sqrt{3} )$ and two degenerate collinear antiferromagnetic (CAFM1, CAFM2) orders (Figs.~\ref{fig:magnetic_order}(d) and~\ref{fig:magnetic_order}(e)) with wave vectors at M$(\pi,\pi/\sqrt{3} )$ and M$^{\prime\prime}$$(\pi,-\pi/\sqrt{3} )$, have no difference in spin configurations except for the direction of the ferromagnetically ordered chains. The spiral order with wave vector at K$(4\pi/3,0)$ and the non-collinear antiferromagnetic (NCAFM) order with wave vector at X$(\pi,0)$  are shown in Figs.~\ref{fig:magnetic_order}(f) and~\ref{fig:magnetic_order}(g) respectively.
Additionally, the one-dimensional antiferromagnetic (1DAFM) order is shown in Fig.~\ref{fig:magnetic_order}(h). In the 1DAFM order, no correlation exists between the horizontal antiferromagnetic chains, resulting in ordering wave vectors uniformly spread over the line connecting M and M$^{\prime}$. The values of equal-time spin structure factor on this line for the 1DAFM order are expected to be much smaller than the peak values for the 2D magnetically ordered states mentioned above~\cite{reuther2011}.

\section{Results}\label{sec:results}

\subsection{Phase diagram}\label{subsec:phase_diagram}
\begin{figure}[htb]
    \centering
    \includegraphics[width=1\linewidth]{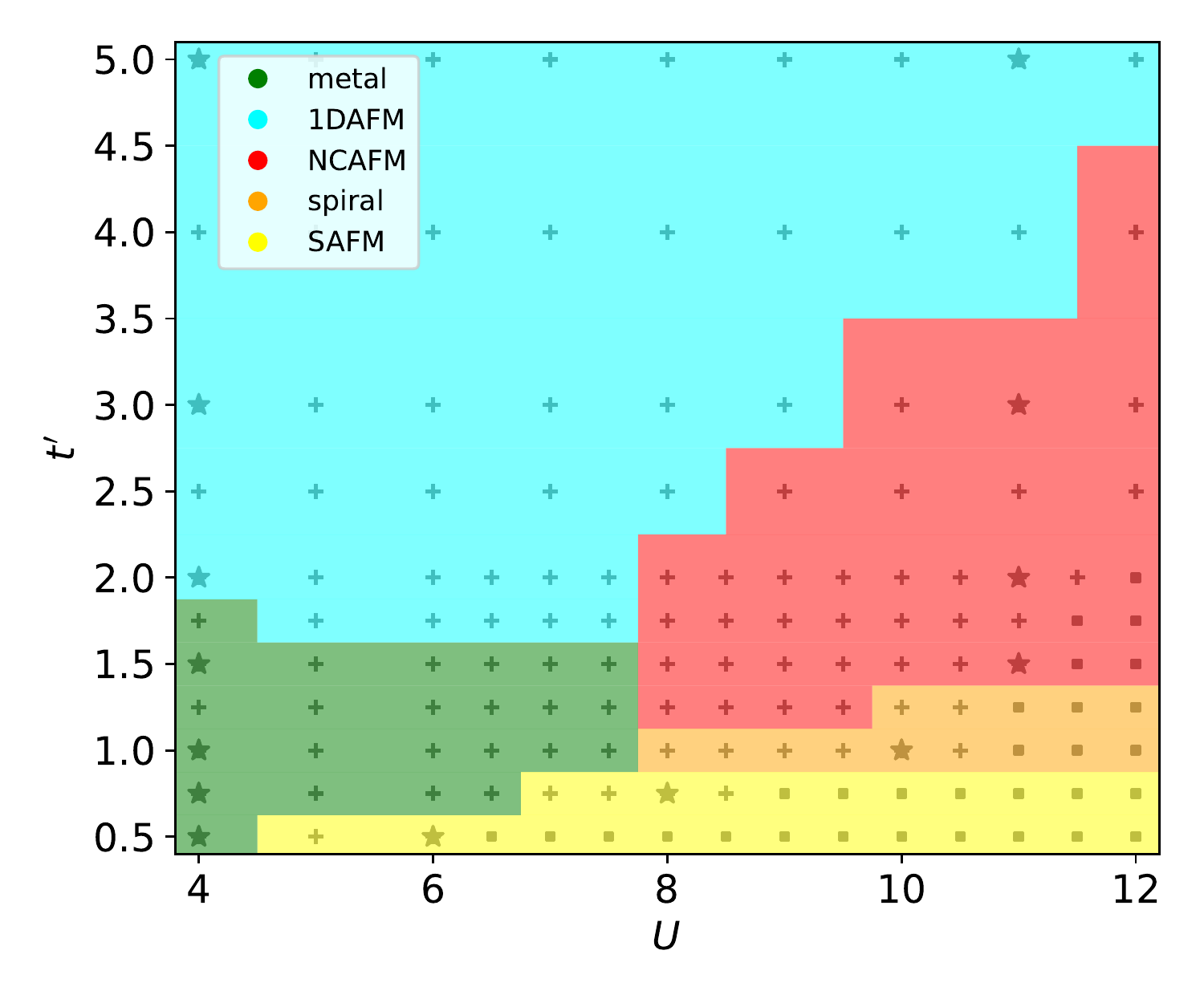}
    \caption{Phase diagram of the anisotropic triangular lattice Hubbard model. Green, cyan, red, orange, and yellow regions correspond to metal, one-dimensional antiferromagnetic (1DAFM), non-collinear antiferromagnetic (NCAFM), spiral, and square-lattice antiferromagnetic (SAFM) orders respectively. Plus markers: places with converged results. Square markers: LDFA fails to converge. Star markers: data points shown in Sec~\ref{subsec:magnetic_phases}. All the points are assigned to the closest commensurate orders or the metallic phase.}
    \label{fig:phase_diagram}
\end{figure}

The phase diagram of the anisotropic triangular lattice Hubbard model, as identified by LDFA, is shown in Fig.~\ref{fig:phase_diagram}. At $4\leq U \leq 6$, the metallic phase dominates around the isotropic lattice limit and transforms into the SAFM order when $t^{\prime}$ is decreased and into the 1DAFM order when $t^{\prime}$ is increased. At $6\leq U \leq 10$, the metal-insulator transitions from the metallic phase to SAFM, spiral, and NCAFM orders emerge at different $t^{\prime}$ due to the increase of $U$. A transition from 1DAFM to CAFM also begins in this region with larger $t^{\prime}$. At $10 \leq U \leq 12$, different magnetic phases appear in the order of SAFM, spiral, NCAFM, and 1DAFM orders upon the increase of $t^{\prime}$ going from the square lattice limit to the isotropic lattice limit to the limit of decoupled chains.  

A $24\times 24$ lattice is used to scan this phase diagram. The physical quantities obtained from the $24\times 24$ lattice are found to have no qualitative difference from the ones obtained from a $48\times 48$ lattice which have higher resolution and are therefore shown in the remainder of the paper.
To identify the different phases in Fig.~\ref{fig:phase_diagram}, we use two quantities that reflect the properties of the system in the charge sector and in the spin sector respectively: the spectral weight at the Fermi energy, $A(\nu=0)=-\operatorname{Im}G(\mathrm{i} 0^{+})/\pi$, and the static spin susceptibility, $\chi_{\text{s}}(\boldsymbol{q}, \omega_n=0)$. 

We first identify the 2D magnetically ordered states according to the absolute values of the static spin susceptibility at the possible magnetic wave vectors mentioned in Sec~\ref{subsection:magnetic_orders}, i.e., the K, M, M$^{\prime}$, M$^{\prime\prime}$, X points shown in Fig.~\ref{fig:magnetic_order}(a). If the largest absolute value at those points is larger than $1$, then we identify the system as belonging to the corresponding magnetically ordered state. As we will show in Sec.~\ref{subsec:magnetic_phases}, incommensurate magnetic orders where the ordering wave vectors do not belong to those points, and a coexisting phase where the features for the 1DAFM and NCAFM orders coexist, are also observed. For simplicity, we attribute them as the closest commensurate magnetic orders in Fig.~\ref{fig:phase_diagram}.

To further classify the remainder of the phase diagram, we define that a point where the spectral weight at the Fermi energy is larger than 0.1 lies in a metallic phase, and otherwise lies in a phase for the 1DAFM order. This way of identification of the metallic phase is consistent with the above method used for the 2D magnetically ordered states. Note that using other threshold values for the static spin susceptibility and the spectral weight instead of $1$ and $0.1$ will generate slightly different phase diagrams but the overall shape of the phase diagram remains similar for those values.

Although our phase diagram does not give precise phase boundaries, our results can reflect the rich phases of the anisotropic triangular lattice systems. A precise determination of the location of the boundaries of phases in the ground state would need additional careful studies of the temperature dependence and of corrections to LDFA which is beyond our current focus. Note that we use the terminology ``1DAFM order/state" in this work to refer to the state that has a straight line feature connecting M and M$''$ in the equal-time spin structure factor, such as the case of the uncorrelated antiferromagnetically ordered chains illustrated in Sec.~II~D. In fact, the quantum state with a such 1D feature found by the DF method is the previously identified 1D QSL which is not a magnetically ordered state~\cite{weng2006,hayashi2007, hauke2011,hauke2013, reuther2011, yunoki2006,heidarian2009,ghorbani2016,castells-graells2019, gonzalez2020, yamada2014a, tocchio2014, szasz2021}.

\subsection{Discussion of previous studies}\label{subsec:previous_studies}

Before discussing the details of our results, we review previous numerical studies on related systems. Both the Hubbard model and the Heisenberg model are known to have a rich ground state phase diagram on the anisotropic triangular lattice. In the discussion of the Heisenberg model, we will focus on $J-J^{\prime}$ model where $J$ is used to denote the nearest-neighbor exchange interaction along two directions and $J'$ the one along the remaining direction. We will not discuss other possible interactions in the spin model such as in-plane isotropy, next-nearest-neighbor exchange, and ring exchange that may or may not be related to the Hubbard model at intermediate $U$~\cite{yang2010}.

In the classical ground state solution of the anisotropic triangular lattice Heisenberg model, the next nearest neighbor chains in the vertical direction are ferromagnetically correlated. The ordering wave vector is given by $(2\theta,0)$ where $\theta=\cos^{-1}[-J/(2J^{\prime})]$ for $0\leq J/J^{\prime}\leq 2$ and $\theta=\pi$ for $J/J^{\prime}>2$~\cite{weihong1999,merino1999}. The real space spin configuration is illustrated in Fig.~\ref{fig:magnetic_order}(b), where $\theta$ is the angle between two nearest neighbor spins in the two diagonal directions while $2\theta$ is the angle between two nearest neighbor spins in the horizontal direction. 
When $J/J^{\prime}$ increases from $0$ to $1$ to infinity, the system undergoes the 1D to 2D transition from NCAFM to spiral to SAFM orders with the ordering wave vector changing smoothly from X to K to M$^{\prime}$. (Note that one of the equivalent points of M$^{\prime}$ is located at $(2\pi,0)$ so there is no jump in the ordering wave vector during the transition.) For the NCAFM order in this classical solution, the spins on the nearest neighbor chains in the vertical direction are orthogonal to each other since $\theta=\pi/2$. Note that a non-orthogonal configuration gives the same energy in the classical solution. Further, the peak at X in the equal-time spin structure factor is only related to the fact that the next nearest neighbor antiferromagnetic chains are ferromagnetically correlated in the vertical direction and are not related to the relative configuration between the nearest neighbor chains. Therefore, when the real-space illustration for the NCAFM order is presented in Fig.~\ref{fig:magnetic_order}(g), we intentionally show a non-orthogonal configuration to leave some ambiguity.

Unlike the classical case, the quantum solution of the Heisenberg model is complicated and is still under debate. In the square lattice limit ($J^{\prime}/J=0$) and the isotropic lattice ($J^{\prime}/J=1$), the SAFM order~\cite{anderson1952,chakravarty1989,manousakis1991} and the spiral order~\cite{huse1988,jolicoeur1989,bernu1992,bernu1994,capriotti1999} are still the ground states, as in the classical problem. The incommensurate orders between the spiral order and SAFM order are also observed~\cite{weihong1999,reuther2011,harada2012,ghorbani2016,gonzalez2020}. However, the transition point between the SAFM order and the spiral/incommensurate orders are found to be shifted above the classical value $J^{\prime}/J=0.5$~\cite{weihong1999,chung2001,weng2006,weichselbaum2011,reuther2011,hauke2011,hauke2013,ghorbani2016,castells-graells2019,gonzalez2020,gonzalez2022}. Furthermore, a dimerized phase is proposed by series expansion (SE)~\cite{weihong1999} and a putative QSL phase is proposed by modified spin wave theory (MSWT)~\cite{hauke2011,hauke2013}, variational Monte Carlo (VMC)~\cite{ghorbani2016}, exact diagonalization (ED)~\cite{castells-graells2019}, and high-temperature SE~\cite{gonzalez2022}  in the region between the SAFM order and the spiral order. Going from the isotropic lattice ($J^{\prime}/J=1$) to the limit of decoupled chains ($J^{\prime}/J=\infty$), the phase diagram becomes more involved, especially between the limit of decoupled chains and the region where the incommensurate orders are found~\cite{weihong1999,bocquet2001,pardini2008,heidarian2009,ghamari2011,reuther2011,thesberg2014,ghorbani2016,gonzalez2020}. While the Tomonaga-Luttinger physics in the 1D antiferromagnetic Heisenberg chain has been clear for some time~\cite{bethe1931,descloizeaux1962,endoh1974,tennant1993}, there is some controversy in the 1D-2D crossover region. On one hand, a gapless QSL, which corresponds to the 1DAFM order in our notation, is expected to occur as a remnant of the Tomonaga-Luttinger liquid in the crossover region and is expected to occupy a large portion of the phase diagram due to dimensional reduction by frustration and quantum fluctuation~\cite{powell2017}. This gapless QSL, along with a strong 1D feature, is confirmed by density matrix renormalization group (DMRG)~\cite{weng2006}, resonating valence bond mean field theory~\cite{hayashi2007}, MSWT~\cite{hauke2011,hauke2013}, functional renormalization group (FRG)~\cite{reuther2011}, VMC~\cite{yunoki2006,heidarian2009,ghorbani2016}, ED~\cite{castells-graells2019}, and Schwinger boson theory~\cite{gonzalez2020}. On the other hand, an unexpected CAFM order, which does not have a classical counterpart, is purposed to be stabilized in the crossover region by renormalization group (RG) studies~\cite{starykh2007,starykh2010} and is later supported by SE~\cite{pardini2008}, coupled cluster method~\cite{bishop2009}, RG~\cite{ghamari2011}, and ED~\cite{thesberg2014}. Besides, DMRG~\cite{weichselbaum2011} and ED~\cite{thesberg2014} studies point out that incommensurate correlations may exist in the entire region going from the square lattice limit to the limit of decoupled chains. In addition, other QSL states besides the 1D QSL are also proposed in the crossover region~\cite{alicea2005,yunoki2006,heidarian2009,hauke2011,castells-graells2019}. 

For the Hubbard model at half-filling, a metallic phase at small $U$ and intermediate anisotropy is expected to be connected with the isotropic non-interacting triangular lattice. The possibility of a superconducting phase in the nearby region is also proposed~\cite{kyung2006,sahebsara2006,watanabe2006} but remains controversial~\cite{clay2008,watanabe2008,tocchio2013,laubach2015}. Recent DMRG studies find that a Luther-Emery liquid (LEL)~\cite{luther1974} could be the true ground state in this region~\cite{gannot2020,szasz2020,szasz2021}. Another recent DMRG study also points out that the superconducting phase only exists with large doping at intermediate $U$~\cite{zhu2022}. Approaching the square lattice limit at small $U$, a metal-insulator transition (MIT) is expected since the SAFM order is known to be the ground state of the square lattice Hubbard model for all $U$~\cite{qin2022}. Approaching the limit of decoupled chains at small $U$, a MIT is reported from variational cluster approximation (VCA)~\cite{yamada2014a} and VMC~\cite{tocchio2014} studies while recent DMRG work also shows the possibility of transiting from metal or LEL to 1D metal~\cite{szasz2021}. Besides the two MITs just mentioned, another interaction-driven MIT around the isotropic lattice limit upon the increasing of $U$ is heavily investigated~\cite{morita2002,kyung2006,aryanpour2006,koretsune2007,clay2008,sahebsara2008,ohashi2008,lee2008,watanabe2008,tocchio2009,yoshioka2009,galanakis2009,kokalj2013,tocchio2013,li2014,yamada2014a,yamada2014,tocchio2014,dang2015,mishmash2015,laubach2015,goto2016,acheche2016,shirakawa2017,misumi2017,szasz2020,li2020,wietek2021,szasz2021,downey2022}. Many studies further propose a QSL or a non-magnetic insulating (NMI) phase existing between the metallic region at small $U$ and the ordered phase at large $U$, which makes a Mott transition between the metallic phase and the QSL or NMI phase possible~\cite{morita2002,kyung2006,koretsune2007,clay2008,sahebsara2008,yoshioka2009,antipov2011,li2014,laubach2015,mishmash2015,shirakawa2017,misumi2017,szasz2020,szasz2021,wietek2021}. Recent studies show evidence of chiral order in the intermediate $U$ region, leading to a conclusion of chiral QSL~\cite{szasz2020,chen2022,cookmeyer2021,szasz2021,wietek2021,zhu2022,kadow2022,zampronio2022a}. At large $U$, similar to the Heisenberg model, the phases at $t^{\prime}/t<1$ are more or less clear with an observation that the transition line between the SAFM order and the spiral/incommensurate orders tend to move above the classical value $t^{\prime}/t=\sqrt{2}/2$~\cite{morita2002,sahebsara2006,watanabe2008,tocchio2013,yamada2014,laubach2015,goto2016,szasz2021}. The proposal of a QSL or a NMI phase existing between the SAFM and spiral orders is also aroused by VMC~\cite{tocchio2009,tocchio2013}, and VCA~\cite{yamada2014}. For $t^{\prime}/t>1$, the phase diagram is as unclear as the corresponding part of the Heisenberg model. VCA~\cite{yamada2014,yamada2014a}, VMC~\cite{tocchio2014},  cluster dynamical mean-field theory~\cite{acheche2016}, and DMRG~\cite{szasz2021} studies find a CAFM order above the spiral/incommensurate orders. The 1DAFM order, i.e., the 1D QSL phase, is also found in a similar region by VCA~\cite{yamada2014a}, VMC~\cite{tocchio2014}, and DMRG~\cite{szasz2021}. Note that the CAFM order and 1DAFM order are found to belong to different regions of the phase diagram in Ref.~\cite{yamada2014a, tocchio2014} but they were found in different DMRG simulation configurations in Ref.~\cite{szasz2021}. The additional QSL phases in the 1D-2D crossover region besides the 1D QSL are also found in some configurations in Ref.~\cite{szasz2021}.

\subsection{Magnetic phases}\label{subsec:magnetic_phases}
\begin{figure}[h!]
    \centering
    \includegraphics[width=0.97\linewidth]{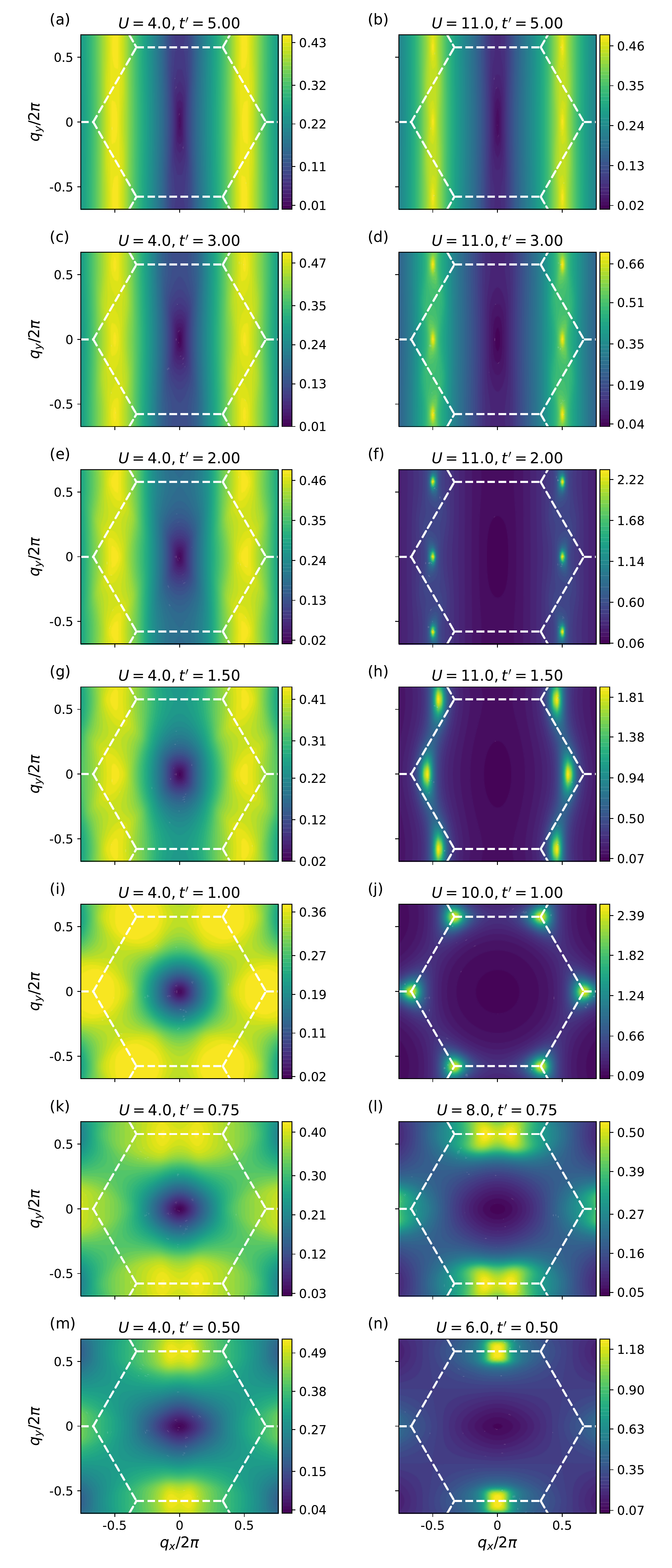}
    \caption{Equal-time spin structure factor $\mathcal{S}^{z z}_{0}(\boldsymbol{q})$ at small $U$ (left column) and large $U$ (right column) with different $t^{\prime}$ ranging from $0.5$ (bottom row) to $5$ (top row). $U$ and $t^{\prime}$ are as indicated. The Brillouin zone is shown by the white dashed lines.}
    \label{fig:ssf}
\end{figure}

Our results for the equal-time spin structure factor as a function of $t ^{\prime}$ at small $U$ and  large $U$ are given in Fig.~\ref{fig:ssf}.

When $t^{\prime}<1$, the SAFM order is observed in a large portion of the parameter space admitting the competition with the spiral order when approaching the isotropic lattice limit. As shown in Figs.~\ref{fig:ssf}(l) and~\ref{fig:ssf}(n), at relatively large $U$, a peak at M$^{\prime}$ is established in equal-time spin structure factor at $t^{\prime}=0.50$ and is then split into two peaks when an incommensurate order is formed at $t^{\prime}=0.75$. A similar transition with less obvious peaks is also observed in the metallic region as shown in Figs.~\ref{fig:ssf}(k) and~\ref{fig:ssf}(m). 
For the isotropic lattice, the peaks at the six corners of the first Brillouin zone are formed when $U$ is gradually increased, as shown in Figs.~\ref{fig:ssf}(i) and~\ref{fig:ssf}(j). No obvious Mott insulating phase is observed at the temperature studied, which agrees with the previous DF results~\cite{li2014} and a recent multimethod study~\cite{wietek2021}. For the potential spin liquid in the middle of the MIT, the multimethod study shows that related roton-like excitations are developed when the temperature decreases from $T=0.1$~\cite{wietek2021}, which can be inferred from the equal-time spin structure factor at the M point. In a similar region, we find convergence difficulties at half-filling below $T=0.1$. Therefore, the non-magnetic insulator phase is absent on our phase diagram. A different starting point for the DF method may alleviate the convergence problem at low temperature~\cite{antipov2011}.

Approaching the limit of decoupled chains with large $U$,  incommensurate order is observed between $t^{\prime}=1$ and  $t^{\prime} \approx 2$ with the peak value in the equal-time spin structure factor evolving from the K point to the X point as shown in Figs.~\ref{fig:ssf}(j),~\ref{fig:ssf}(h), and~\ref{fig:ssf}(f). At even larger $t^{\prime}$, as shown in  Figs.~\ref{fig:ssf}(b) and~\ref{fig:ssf}(d), the peak at X decreases while a background broadening ranging from M to M$^{\prime}$ emerges and finally becomes comparable to the peak at X. A similar trend is also observed at small $U$ as shown in Figs.~\ref{fig:ssf}(i),~\ref{fig:ssf} (g),~\ref{fig:ssf}(e),~\ref{fig:ssf}(c), and~\ref{fig:ssf}(a), where the broadening 1D feature corresponding to the 1DAFM order appears much early with the increase of $t^{\prime}$. Despite the occurrence of the 1DAFM order, we note that the peak at X seems to be stable and may extend to large $t^{\prime}$ for both small and large $U$.

Our equal-time spin structure factor results at large $U$ clearly show a smooth transition from the NCAFM order to the 1DAFM order, which is remarkably similar to the susceptibility or structure factor shown in the FRG study of the Heisenberg model~\cite{reuther2011} and the ED study of the XX model~\cite{yuste2017}. The unexpected CAFM order proposed in the Heisenberg model~\cite{starykh2007,starykh2010,pardini2008,bishop2009,ghamari2011,thesberg2014} and the Hubbard model~\cite{yamada2014,yamada2014a,tocchio2014,acheche2016,szasz2021} is not found in our calculation. Furthermore, the coexisting state, where both the peak at X and the line going from M to M$^{\prime}$ are visible in the equal-time spin structure factor such as the one shown in Fig.~\ref{fig:ssf}(d), is found to connect the NCAFM order and the 1DAFM order despite that we classify them to the nearest ordered phase in Fig.~\ref{fig:phase_diagram}. To our best knowledge, this coexisting state has not yet been discussed and could be directly related to the controversy in the large $t^{\prime}/t$ or large $J^{\prime}/J$ region. We find some literature indeed providing indirect evidence for this coexisting state. For example, both DMRG~\cite{weichselbaum2011} and ED~\cite{thesberg2014} studies for the Heisenberg model support that the incommensurate correlation exists in the entire region going from the isotropic lattice to the limit of decoupled chains which agrees with the observation here that the peak at X tend to exist at very large $t^{\prime}$. Although the ED study~\cite{thesberg2014} observes a phase transition from the incommensurate order to the CAFM order instead of the NCAFM order as $J^{\prime}$ increases, the energy of the NCAFM order is found to be larger than the CAFM order only by a margin of $\sim 10^{-7}$ at large $J^{\prime}$, and this small energy difference even decreases with system size. Besides, a recent DMRG study on the Hubbard model also finds a similar phase like the coexisting phase found here in the YC6 cylinder~\cite{szasz2021}. The supplemental materials of this work also point out that the 1D QSL phase found by the work may be non-trivial because of the isolated gapless points (probably at $\Gamma$ and X). We note that this is consistent with the picture of the NCAFM order coexisting with the 1DAFM order presented here. Furthermore, a recent VMC work on the Hubbard model find a similar coexisting state of the CAFM order and the 1DAFM order where additional anisotropy is introduced for the hopping terms in the model such that CAFM can be stabilized in the nearby regions~\cite{ido2022}. In the VMC work, this coexisting phase is further identified as a new type of QSL. Lastly, a smooth transition from the 1DAFM order to the SAFM order is also indicated in the auxiliary field quantum Monte Carlo (AFQMC) study of the anisotropic square lattice Hubbard model with additional frustration from the next-nearest neighbor hopping~\cite{raczkowski2013}.

Besides the 1DAFM order and the coexisting state discussed above, the phases slightly above and below the isotropic lattice limit at large $U$ may also not have  counterparts in the classical solution of the Heisenberg model. As shown in Figs.~\ref{fig:ssf}(h) and~\ref{fig:ssf}(l), the amplitude of the equal-time spin structure factor decreases in these two regions and the peaks get blurred. We note that it is possible to relate the characteristics of the equal-time spin structure factor for the coexisting state and the phases shown in Figs.~\ref{fig:ssf}(h) and~\ref{fig:ssf}(l) to QSL phases~\cite{heidarian2009,yuste2017,ido2022}. However, we do not have other reliable methods to classify QSL within the framework of LDFA and we cannot obtain convergence for even larger $U$ to discuss the stability of these phases at the infinite $U$ limit. Therefore, we conservatively classify them as the nearest ordered phases in the phase diagram, Fig.~\ref{fig:phase_diagram}.

\subsection{Lifshitz transitions}\label{subsec:fermi_surface_opening}

\begin{figure}[htb]
    \centering
    \includegraphics[width=1\linewidth]{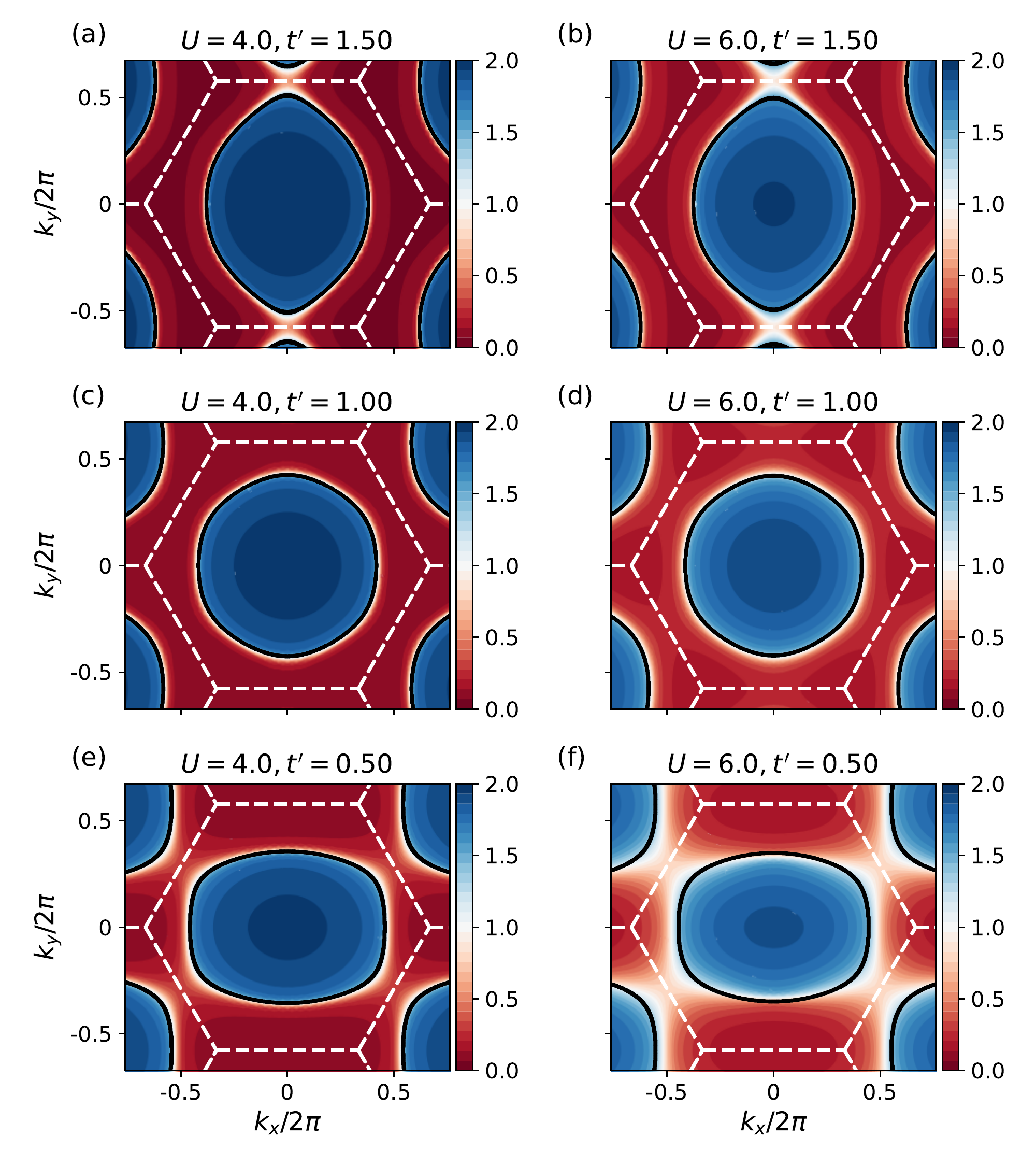}
    \caption{Electron occupation $n(\boldsymbol{k})$ at $U=4$ (left column) and $U=6$ (right column) with different $t^{\prime}$ ranging from $0.5$ (bottom row) to $1.5$ (top row). $U$ and $t^{\prime}$ are as indicated. Black lines show the non-interacting Fermi surfaces. The Brillouin zone is shown by the white dashed lines.}
    \label{fig:density_k}
\end{figure}

For the small $U$ region of the phase diagram, we identify the 1DAFM order from the metallic phase according to the spectral weight at the Fermi energy in Sec.~\ref{subsec:phase_diagram}. In fact, the same boundary between these two phases is also related to the Lifshitz transitions which in our case is the Fermi surface opening at the boundary of the Brillouin zone. Fig.~\ref{fig:density_k} shows the electron occupation $n(\boldsymbol{k})=\sum_{\sigma, \omega_n} \mathrm{e}^{-\mathrm{i} \omega_n 0^{-}} G_{\sigma}(\mathrm{i} \omega_n, \boldsymbol{k})/\beta$ at $U=4$ and $U=6$ with different $t^{\prime}$ ranging from $0.5$ to $1.5$. The black line indicates the non-interacting Fermi surface $\varepsilon(\boldsymbol{k})-\mu=0$. We see that at $U=4$ the Fermi surface (white regions) roughly matches the non-interacting Fermi surface at all $t^{\prime}$ and two Lifshitz transitions inherited from the non-interacting limit occur at $t^{\prime}$ slightly above $1.5$ and below $0.5$. At $U=6$, with the changing of $t^{\prime}$ the Fermi surface may have already opened at $t^{\prime}=0.5$ and $t^{\prime}=1.5$, as indicated by the white regions in Figs.~\ref{fig:density_k}(b) and~\ref{fig:density_k}(f). In the recent DMRG study of the same system, the Fermi surface opening is also observed at very close regions and the trend that the Fermi surface at larger $U$ opens at $t^{\prime}$ closer to $1$ also agrees well~\cite{szasz2021}. Besides, as shown in Fig.~\ref{fig:density_k}, in the presence of interactions the saddle points of Lifshitz transitions for the non-interacting case become broadened white regions and at larger $U$ the Fermi surface encloses effectively a larger region. This phenomenon is well understood in a Landau-type theory~\cite{volovik2017} as flat bands near the conventional Lifshitz transition. Similar phenomena are also found in the isotropic triangular lattice Hubbard model at van Hove filling~\cite{yudin2014} and the doped Kondo lattice model~\cite{peters2015}.

We note that in Ref.~\cite{szasz2021}, although a similar Lifshitz transition is found, the small $U$ and large $t^{\prime}$ region (left top corner of Fig.~\ref{fig:phase_diagram}) is recognized as 1D metal instead of 1DAFM. In our calculation, lower temperature data are needed to distinguish the insulating phase and a correlated metal. This is beyond the scope of the method. We note that the insulating phase is found in VCA~\cite{yamada2014a} and VMC~\cite{tocchio2014} in similar regions, which agrees with the 1DAFM order at large $t^{\prime}$ we found. Besides, a similar transition from a 1D insulating phase to a higher-dimensional metallic phase with antiferromagnetic spin correlations is reported from the AFQMC study of the anisotropic square lattice Hubbard model~\cite{raczkowski2015}.  Therefore, the metallic or insulating nature at the small $U$ and large $t^{\prime}$ region may need further study.

\begin{figure*}[htb]
    \centering
    \includegraphics[width=1\linewidth]{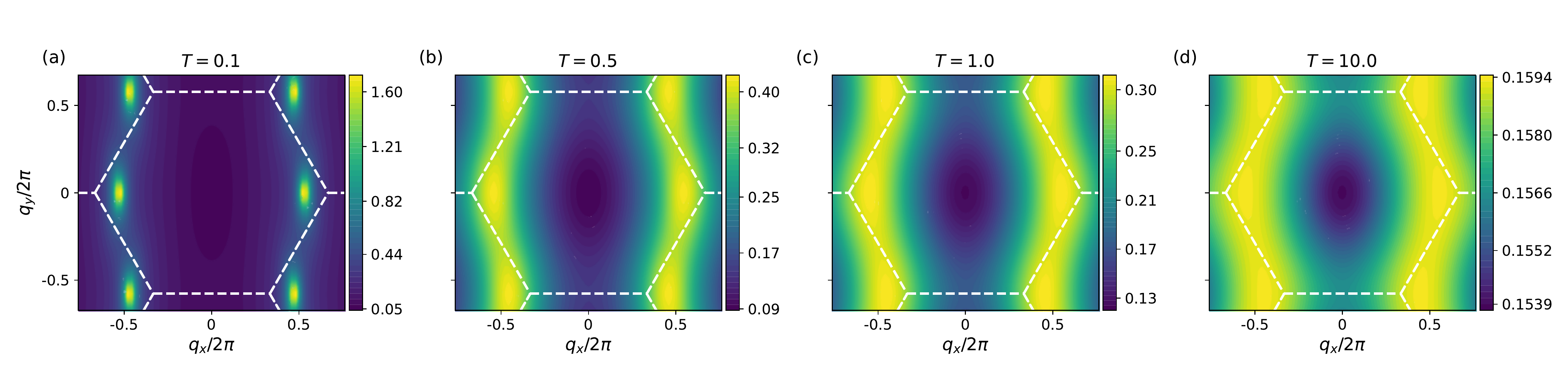}
    \caption{Equal-time spin structure factor $\mathcal{S}^{z z}_{0}(\boldsymbol{q})$ at $U=11$ and $t^{\prime}=1.715$ with different temperatures $T$ ranging from $0.1$ (left) to $10$ (right). $T$ is as indicated. The Brillouin zone is shown by the white dashed lines.}
    \label{fig:ssf_Cs2CuCl4}
\end{figure*}
\begin{figure*}[htb]
    \centering
    \includegraphics[width=1\linewidth]{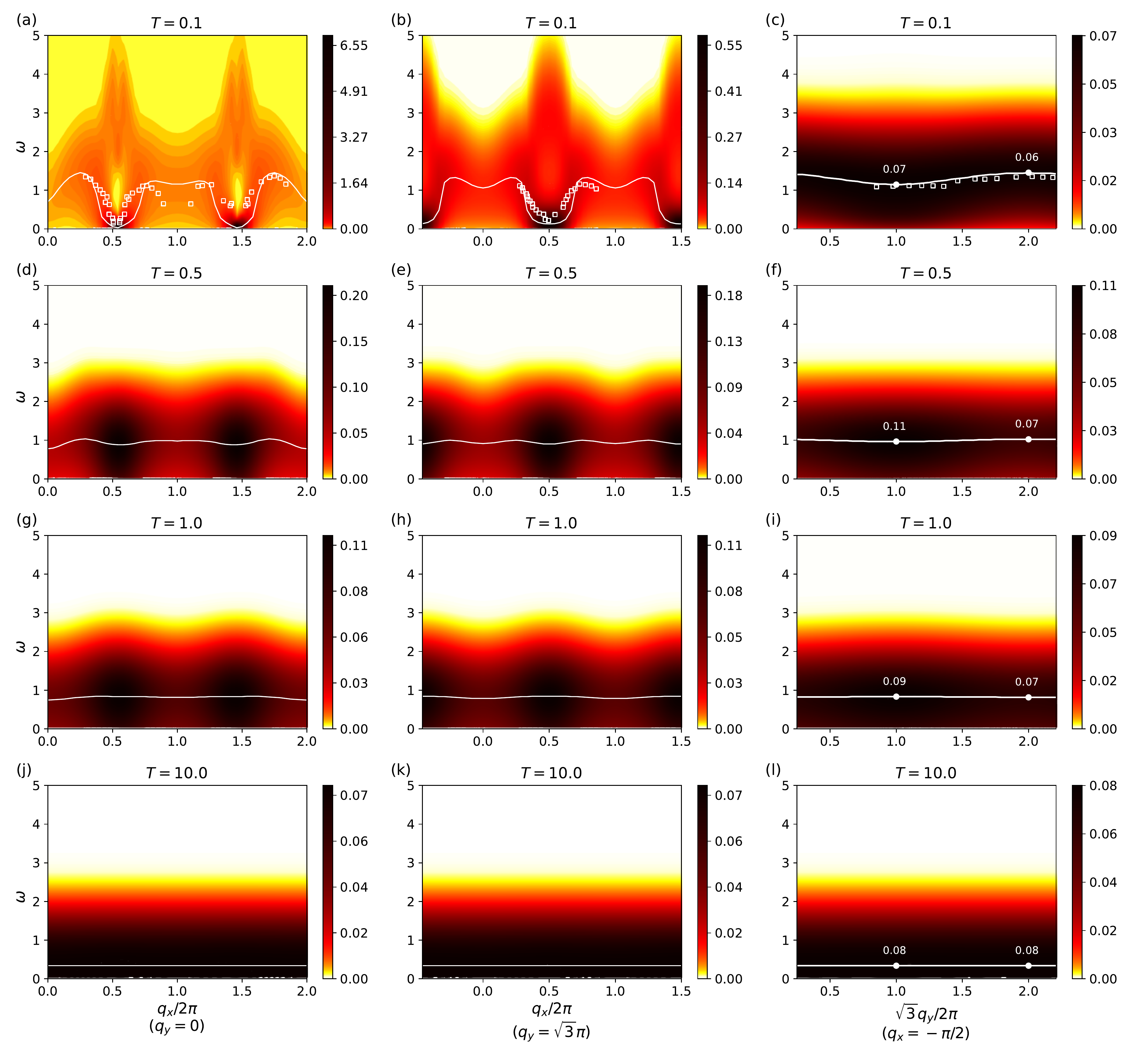}
    \caption{Dynamical spin structure factor $\mathcal{S}^{z z}(\boldsymbol{q},\omega)$ at $U=11$ and $t^{\prime}=1.715$ with different temperatures $T$ ranging from $0.1$ (top row) to $10$ (bottom row). $T$ is as indicated. Left column, middle column, and right column show $\mathcal{S}^{z z}(\boldsymbol{q},\omega)$ at $q_y=0$, $q_y=\sqrt{3} \pi $, and $q_x=-\pi/2$, respectively. White lines shows the locations of peaks for each fixed $\boldsymbol{q}$. White open squares are peak values at $T<0.1$K reported from the neutron scattering experiment with a rescaling by a factor of two~\cite{coldea2003}. Values of peaks at $q_y=2\pi/\sqrt{3} $ and $q_y=4\pi/\sqrt{3} $ are shown in the right column with white solid dots and numbers.}
    \label{fig:dssf_Cs2CuCl4}
\end{figure*}

\subsection{Spin excitation of  \texorpdfstring{$\text{Cs}_2\text{CuCl}_4$}{TEXT}}

The anisotrpic triangular lattice Hubbard model at $t'=1.715$ (estimated from $J/J^{\prime}=0.34$) is believed to describe the physics of Cs$_2$CuCl$_4$. We present the spin structure factors and compare them with the inelastic neutron scattering experiment of Cs$_2$CuCl$_4$~\cite{coldea2003}. Our results overall show good agreement with Ref.~\cite{coldea2003}. Besides, we find a region where a distorted 1D feature and the feature of an incommensurate order coexist in the equal-time spin structure factor. This is similar to the coexisting states of the NCAFM and 1DAFM orders discussed in Sec.~\ref{subsec:magnetic_phases}. 

Fig.~\ref{fig:ssf_Cs2CuCl4} shows the equal-time spin structure factor at $U=11$ and $t^{\prime}=1.715$ with different temperatures. Upon increasing the temperature from $T=0.1$ to $T=1$, the peaks at incommensurate ordering wave vector shown in Fig.~\ref{fig:ssf_Cs2CuCl4}(a) gradually evolve into broadening 1D curves as shown in Fig.~\ref{fig:ssf_Cs2CuCl4}(c). At intermediate temperature $T=0.5$ as shown in Fig.~\ref{fig:ssf_Cs2CuCl4}(b), both the blurred peaks at the incommensurate ordering wave vector and a distorted 1D feature are observed. Such coexistence actually also appears at $T=0.1$ and $T=1$ with either the 1D feature or the peak at incommensurate ordering wave vector much less obvious. At high temperature $T=10$, as shown in Fig.~\ref{fig:ssf_Cs2CuCl4}(d), the structure factor becomes almost featureless considering the minor difference between the maximum and the minimum. The finite temperature transition presented above is consistent with the transition from an incommensurate magnetic order to a spinon excitation continuum to a paramagnetic state observed in the inelastic neutron scattering experiment~\cite{coldea2003}. The coexisting states also appear to be consistent with the dispersion relation given in the experiment.

To further illustrate the spin excitations, we also present the dynamical spin structure factor in Fig.~\ref{fig:dssf_Cs2CuCl4}  with momentum cuts at $q_y=0$, $q_y=\sqrt{3} \pi $, and $q_x=-\pi/2$, following the ones used in Fig.~3 of Ref.~\cite{coldea2003}. The comparison of the spin excitations from the experiment and calculations based on the Heisenberg model are well known. The peak dispersion of neutron scattering is well described by series expansion and $1/S$ expansion~\cite{veillette2005,dalidovich2006,zheng2006,fjaerestad2007}. A perturbative calculation based on 1D spin chains further reveals the excitation continuum and the peak dispersion originate from lower dimension spinons~\cite{kohno2007}.

Due to the ill-conditioned nature of the analytic continuation problem, the details of the dynamical spin structure factor depend on the default model used in maximum entropy calculation and are also sensitive to the error estimation of the calculation which is difficult to obtain in the LDFA calculation. This limits the reliability of the dynamical quantities and thus makes the direct comparison with experimental data difficult. Since fine-tuning the parameters in the analytic continuation process may introduce misleading agreement, we conservatively choose to use the same model and same error estimation for all momentum points $\boldsymbol{q}$ and all temperatures $T$ studied.  This results in anomalies that could be removed by tuning the parameters such as high energy tails around $q_{x}=\pi,3\pi$ in Fig.~\ref{fig:dssf_Cs2CuCl4}(a) and around $q_{x}=\pi$ in Fig.~\ref{fig:dssf_Cs2CuCl4}(b) where the Matsubara data are significantly larger than at other $\boldsymbol{q}$ points.

As shown in the top row of Fig.~\ref{fig:dssf_Cs2CuCl4}, the peak dispersion (white line) at low temperature $T=0.1$ qualitatively matches the peak dispersion (white open square) below $0.1$K given in the neutron scattering experiment~\cite{coldea2003}. 
More precisely, the minimum close to $q_x=\pi$ and $q_x=3\pi$ shown in Fig.~\ref{fig:dssf_Cs2CuCl4}(a), the asymmetric dispersion with respect to $q_x=\pi$ and $q_x=3\pi$ shown in Fig.~\ref{fig:dssf_Cs2CuCl4}(a), the symmetric dispersion with respect to $q_x=\pi$ shown in Fig.~\ref{fig:dssf_Cs2CuCl4}(b), and the asymmetric dispersion with respect to $q_y=\sqrt{3}\pi$ shown in Fig.~\ref{fig:dssf_Cs2CuCl4}(c) all agree with the dispersion relation obtained in the experiment. Note that to highlight this agreement, the experimental results have been scaled by a factor of two. Using the estimate $J=4t^2/U$ and $J=0.128(5)$meV~\cite{coldea2003}, one can show $T=0.1$ used in the calculation corresponds to a physical temperature above $0.1$K but below $T_{N}=0.62$K, the transition temperature from an incommensurate order to excitation continuum reported in Ref.~\cite{coldea2003}. Adjusting temperature $T$ (and/or $U,t'$) therefore may yield a more accurate absolute agreement here. Besides the agreement for dispersion, the peak value at $\boldsymbol{q}=(-\pi/2, 2\pi/\sqrt{3}) $ is found to be larger than the peak value at $\boldsymbol{q}=(-\pi/2, 4\pi/\sqrt{3})$ as indicated by the white solid dots and numbers in Fig.~\ref{fig:dssf_Cs2CuCl4}(c). The ratio of these two peak values is found to be about $4$ in the experiment (see scan E, F in Fig.~5 of Ref.~\cite{coldea2003}) and is well explained by the perturbative study~\cite{kohno2007}. Although the ratio found in our calculation is much smaller than the experiment value, we find the feature of a larger peak value occurring at  $\boldsymbol{q}=(-\pi/2, 2\pi/\sqrt{3}) $ is robust upon different default models and error estimation.  

When temperature $T$ is further increased, the dispersion tends to become flat as shown in the second to the fourth row of Fig.~\ref{fig:dssf_Cs2CuCl4} which signatures the features of the incommensurate order at lower temperature gradually disappear. The ratio of peak values at  $\boldsymbol{q}=(-\pi/2, 2\pi/\sqrt{3})$  and $\boldsymbol{q}=(-\pi/2, 4\pi/\sqrt{3})$ becomes larger at $T=0.5$ and then decreases, which agrees with the picture that spinon excitations are enhanced at intermediate temperature and the paramagnetic state appears at higher temperature~\cite{coldea2003}.

\section{Conclusion}\label{sec:conclusion}

We have presented a comprehensive study of the phase diagram of the anisotropic triangular lattice Hubbard model as a function of $t^{\prime}$ and $U$ at half-filling.

The ladder dual fermion approximation used here captures all local correlations while treat non-local correlations perturbatively. It provides a fine k-space resolution of the susceptibility and is able to resolve spin fluctuations without an \textit{a priori} assumption of possible orders.

We found a rich phase diagram with metallic, SAFM, spiral, NCAFM, and 1DAFM phases. While many aspects of the phase diagram have been found before, we resolve the discrepancies between previously published results. More precisely, the disputed CAFM order (see Sec.~\ref{subsec:previous_studies}) near the limit of decoupled chains is not found by our method. Instead, the NCAFM order and the coexisting phase of the 1DAFM and NCAFM orders are found to be dominant in the corresponding region. Some indirect support in previous studies related to the newly found coexisting phase is discussed in Sec.~\ref{subsec:magnetic_phases}. We also investigated the Lifshitz transition on the anisotropic triangular lattice in the presence of interaction. 

Our study investigated the physics of the model for parameters relevant to Cs$_2$CuCl$_4$ in more detail. We found a transition from an incommensurate magnetic order to a spinon excitation continuum to a paramagnetic state upon the increase of temperature, which is in agreement with the inelastic neutron scattering experiment~\cite{coldea2003}. Despite the limitation of analytic continuation, the dynamical spin structure factor at $U=11$ and $t'=1.715$ shows a dispersion relation consistent with results obtained in Ref.~\cite{coldea2003}.

\section*{Acknowledgement}
We thank Wei-Ting Lin and André Erpenbeck for helpful discussions. 
This work was supported by NSF DMR 2001465. This work used Expanse at SDSC through allocation DMR130036 from the Extreme Science and Engineering Discovery Environment (XSEDE), which was supported by National Science Foundation grant number \#1548562.

\appendix

\section{Local density of states for the non-interacting anisotropic triangular lattice}\label{app:DOS}
The local density of states, $\rho(\nu)=-\operatorname{Im}G^{0}(\nu+\mathrm{i} 0^{+})/\pi$, of the non-interacting anisotropic triangular lattice is used as the input of DMFT (see Ref~\cite{horiguchi1991} for the expression of the non-interacting Green's function $G^{0}$). It can be expressed analytically in terms of the complete elliptic integral of the first kind $K(m)=\int_{0}^{\pi / 2} d \phi \left[1-m \sin^{2}\phi\right]^{-1 / 2} $ as:
\begin{equation}
    \rho(\nu; t, t^{\prime}) =  \frac{1}{\pi^2 t^{\prime} \sqrt{z_0 } } K\left( \frac{z_1 }{z_0 } \right),
\end{equation}
\begin{equation}
z_0 = \left\{ 
\begin{aligned}
    &q  \qquad   &0<p\leq q\\
    &p  \qquad   &0<q<p\\
    &p-q \qquad &q<0,
\end{aligned}
\right.
\end{equation}
\begin{equation}
    z_1 = \left\{ 
    \begin{aligned}
        &q-p  \qquad &   0<p\leq q \\
        &p-q  \qquad &  0<q<p \\
        &p \qquad &q<0,
    \end{aligned}
    \right.
    \end{equation}
where we define two dimensionless parameters $u=t/t^{\prime}\geq 0$ and $E=\nu/t$ along with the following notation:
\begin{equation}
    \begin{aligned}
        r& \equiv u\sqrt{u^2-Eu+2},   \\    
        p&\equiv 4r,\\
        q &\equiv  \frac{(r - u^2)^2 (r^2 - 4 u^2 + 2 r u^2 + u^4)}{4 u^4}.\\
    \end{aligned}
\end{equation}
The range of $\nu$ is indicated by 
\begin{equation}
    \begin{aligned}
        &-4-\frac{2}{u}\leq E \leq u+\frac{2}{u} \qquad &  0<u\leq2,\\
        &-4-\frac{2}{u}\leq E \leq 4-\frac{2}{u} \qquad & u>2.
    \end{aligned}
\end{equation}

\bibliographystyle{apsrev4-2}
\bibliography{ref}

\end{document}